# Recent Advancements in the Development of Two-dimensional Transition Metal Dichalcogenides (TMDs) and their potential application

Mitesh B. Solanki [a*], Margi Jani [b]

[a] *Department of Applied Science, Parul Institute of Technology, Parul University, Vadodara, Gujarat-391760, India.*

[b] *Department of Science, Pandit Deendayal Energy University, Gandhinagar, Gujarat, India.*
Author of Correspondence: miteshkumar.solanki29907@paruluniversity.ac.in;

**Abstract**

This article explores the recent advancements in atomically thin two-dimensional transition metal dichalcogenides (2D TMDs) and their potential applications in various fields, including nanoelectronics, photonics, sensing, energy storage, and optoelectronics. Specifically, the focus is on TMDs such as $MoS_2$, $WS_2$, $MoSe_2$, and $WSe_2$, promising for next-generation electronics and optoelectronics devices based on ultra-thin atomic layers. One of the main challenges in utilising TMDs for practical applications is the scalable production of defect-free materials on desired substrates. However, innovative growth strategies have been developed to address this issue and meet the growing demand for high-quality and controllable TMD materials. These strategies are compatible with conventional and unconventional substrates, opening up new possibilities for practical implementation. Furthermore, the article highlights the development of novel 2D TMDs with unique functionalities and remarkable chemistry. These advancements contribute to expanding the range of applications and capabilities of TMD materials, pushing the boundaries of what can be achieved with these ultra-thin layers. In addition to electronics, the article delves into the significant efforts dedicated to exploring the potential of 2D TMDs in energy and sensor applications. These materials have shown promising characteristics for energy storage and have been extensively studied for their sensing capabilities, showcasing their versatility and potential impact in these fields. This article provides a comprehensive overview of the recent progress in 2D TMDs, emphasising their applications in electronics, optoelectronics, energy, and sensing. The continuous research and development in this area is promising for advancing these materials and their integration into practical devices and systems.

1. **Introduction**

Graphene's remarkable success has spurred an astonishing surge in developing additional two-dimensional (2D) materials that exhibit exceptional properties when formed into atomic sheets. Surprisingly, the collection of 2D materials grows yearly, comprising over 150 uniquely layered materials that can be readily separated into ultra-thin subnanometer sheets[1]–[3]. Examples of these materials include 2D transition metal dichalcogenides (TMDs) such as molybdenum disulfide ($MoS_2$), molybdenum diselenide ($MoSe_2$), tungsten disulfide ($WS_2$), and tungsten diselenide ($WSe_2$), as well as hexagonal boron nitride (h-BN), borophene (2D boron), silicene (2D silicon), germanene (2D germanium), and MXenes[4], [5]. Figure 1 depicts a chronological list of publications on 2D materials, highlighting the increasing research focus on TMDs [6]. Based on their chemical compositions and structural arrangements, these atomically thin 2D materials can be classified as metallic, semi-metallic, or non-metallic. Among these materials, semi-metallic, semiconductor, insulator, or superconductor TMDs have attracted significant research attention as the earliest successors of graphene[7], [8]. They possess similar thinness, transparency, and flexibility as graphene[9] [10]. Unlike graphene, many 2D TMDs exhibit semiconductor behaviour and hold the potential to enable ultra-small, low-power transistors that surpass the capabilities of current silicon-based transistors, which face challenges in scaling down[11], [12]. Furthermore, TMDs can be deposited onto flexible substrates, demonstrating resilience against stress and strain while offering bandgaps in the visible-to-near-infrared range, high carrier mobility, and on/off ratios comparable to silicon[13], [14] [11]. This combination of characteristics positions TMDs as promising candidates for future electronic devices, particularly in flexible electronics.

The transition from bulk materials to monolayers introduces quantum confinement and surface effects. This leads to distinct electrical and optical properties in two-dimensional (2D) transition metal dichalcogenides (TMDs). These TMDs exhibit controllable bandgaps, high photoluminescence (PL), and significant exciton binding energy, making them highly appealing for various optoelectronic devices such as solar cells, photodetectors, light-emitting diodes, and phototransistors [15]–[19]. For instance, MoS2, with its direct bandgap of 1.8 eV, robust mobility of 700 cm2 V1 s1, a high current on/off ratio ranging from 107 to 108, substantial optical absorption (107 m1 in the visible region), and strong photoluminescence originating from the direct bandgap in its monolayer form, has garnered significant attention for applications in electronics and optoelectronics[20]. These unique features position 2D TMDs as promising candidates for advancing the field of optoelectronics and driving technological innovation[21].

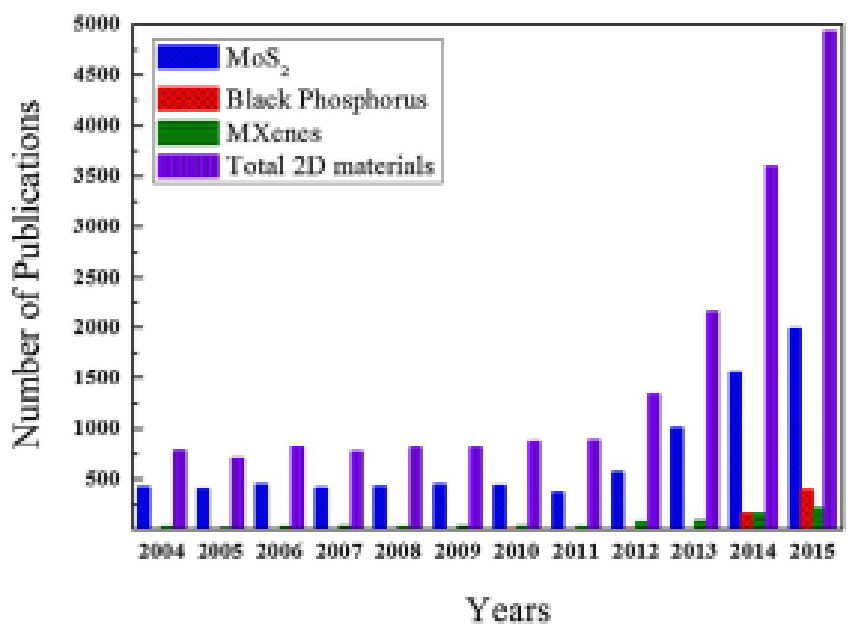

Figure 1 illustrates the publication trends of various 2D transition metal dichalcogenides (TMDs), including MoS2, MoSe2, black phosphorus, MXenes, and the overall total of 2D TMDs. The data presented in this chart covers the period from 2005 to 2016 and was obtained through searches conducted using SciFinder Scholar (https://scifinder.cas.org) and the American Chemical Society database (https://www.acs.org/content/acs/en.html) as of August 10th, 2016.

Van der Waals (vdW) gaps between adjacent layers and the large specific surface area resulting from their sheet-like structures are distinguishing characteristics that make 2D transition metal dichalcogenides (TMDs) highly appealing for capacitive energy storage, as well as for applications in supercapacitors, batteries, and sensing devices[22], [23]. The high surface-to-volume ratio of 2D TMDs contributes to enhanced sensitivity, selectivity, and reduced power consumption in TMD-based sensors. Unlike digital sensors, TMDs-based sensors do not rely on physical gates to selectively respond to targeted gases or molecules, making them suitable for gas, chemical, and biosensing applications[24], [25]. Moreover, the weak interlayer bonding in 2D TMDs enables their separation and stacking with other TMDs to create a diverse range of van der Waals heterostructures without the need for lattice matching[26], [27]. These vertically stacked heterostructures can achieve functionalities and superior qualities that would otherwise be unattainable [21], [28]. This opens up avenues for the fabrication of various electronic and optoelectronic devices, including tunnelling transistors, barristers, photodetectors, LEDs, and flexible electronics, by utilising the unique properties of these vdW heterostructures, such as band alignment, tunnelling transport, and interlayer solid coupling [27], [29]. Figure 2 showcases a variety of devices made from 2D TMDs, highlighting their distinctive physical, chemical, and optoelectronic features [30]–[34].

Despite the unique properties exhibited by two-dimensional (2D) transition metal dichalcogenides (TMDs) that set them apart from traditional bulk materials or thin films, achieving large-scale, defect-free atomic layers with precise thickness control on desired substrates remains challenging. Mechanical exfoliation is the most advanced technique for obtaining high-quality TMD monolayers, but it lacks scalability [35], [36]. Chemical vapour deposition (CVD) presents a promising alternative due to its scalability and morphological control capabilities [37]. Recent progress in CVD has led to significant improvements in the quality of TMD layers produced through this method. Metal-organic CVD (MOCVD) and atomic layer deposition (ALD) are other potential approaches for fabricating wafer-scale, high-quality TMD films [38], [39].

The study of 2D materials is a rapidly growing field, and exploring new materials beyond graphene extends beyond TMDs. Silicene and phosphorene, for instance, have emerged as strong contenders in the expanding landscape of 2D materials[38], [40]. While theoretical investigations have shed light on the fundamental characteristics of these novel materials, their experimental exploration is still in its early stages due to stability challenges[41].

This study focuses on recent advancements in synthesising large-scale, defect-free 2D TMDs. Additionally, we highlight the progress made in understanding the electrical, optoelectronic, and electrochemical properties of newly investigated TMDs, focusing on rational designs and innovative structures for future applications in electronics, sensors, and energy storage[42]. Furthermore, we discuss recent developments in emerging 2D materials, such as silicene and phosphorene. Given their fascinating range of features and potential for applications in emerging technologies, TMDs are expected to remain a prominent research topic [43].

## 2. Physical characteristics and crystal structure

Two-dimensional transition metal dichalcogenides (TMDs) consist of a transition metal (M) layer sandwiched between two atomic layers of chalcogens (X)[6]. The structures of 2D TMDs can be categorised into trigonal prismatic (hexagonal, H) and octahedral (tetragonal, T). Their distorted phase (T0) depends on the arrangement of atoms, as depicted in Fig. 3a. In most cases, multilayer TMDs follow the atomic ratio of one transition metal to two chalcogen atoms, forming MX2 compounds, except for a few exceptions like 2:3 quintuple layers (M2X3) [40], [44]and 1:1 metal chalcogenides (MX) [45]. In the H-phase, each metal atom forms two tetrahedra extending in the +z and z directions, resulting in a hexagonal symmetry when observed from the top view (Fig. 3a). Therefore, the arrangement of chalcogen-metal-chalcogen along the z-direction is considered a single layer. The weak van der Waals interactions between layers (chalcogen-chalcogen) allow for mechanical exfoliation, producing single-layer flakes from bulk TMDs[46]. The T-phase exhibits a trigonal chalcogen layer on top and a 180-degree rotating structure (trigonal antiprism) at the bottom. This leads to a hexagonal arrangement of chalcogen atoms in the top view. The metal atoms undergo further deformation or dimerisation in one direction, resulting in a change in the atomic displacement of chalcogen atoms along the z-axis[47], [48].

Despite graphene's impressive electron mobility (e.g., 15,000 cm2 V1 s1 at ambient temperature), its lack of a bandgap restricts its application as an active component in field-effect transistors (FETs) [49]. Numerous efforts have been made to open the bandgap of graphene through methods such as nanoribbon formation, AB-stacked bilayer graphene, and chemical doping. However, these approaches have achieved limited success, typically resulting in bandgaps of up to 200 meV [50]–[52].

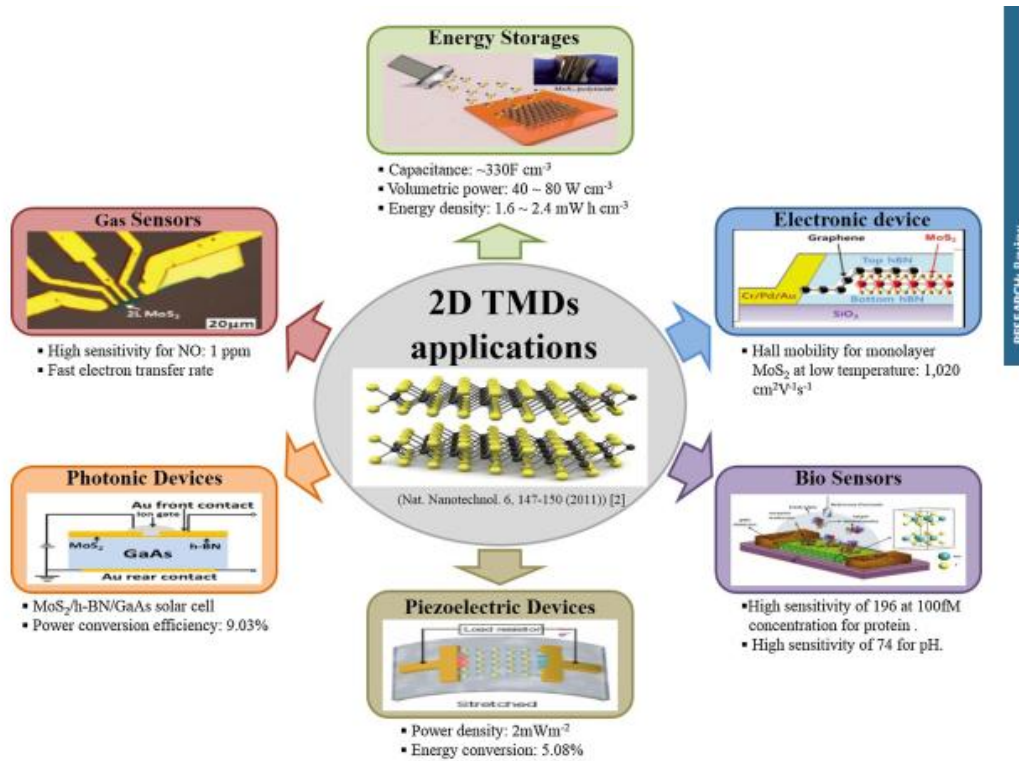

Figure 2 displays the utilisation of two-dimensional transition metal dichalcogenides (TMDs) in various electrical, optoelectronic, and energy devices. The figure includes references[30][31]–[34],

Addressing the limited bandgap in two-dimensional transition metal dichalcogenides (TMDs) has been a challenging problem driving their development. Figure 3b illustrates the wide range of bandgaps exhibited by different 2D TMDs, spanning the visible and infrared regions[53]. While a few exceptions like GaSe and ReS2 exist, most semiconducting TMDs demonstrate a direct bandgap in monolayer form and an indirect bandgap in bulk form [23],

[54]. Monolayer dichalcogenides, such as MoS2 (1.8 eV), MoSe2 (1.5 eV), (2H)-MoTe2 (1.1 eV), WS2 (2.1 eV), and WSe2 (1.7 eV), exhibit direct bandgaps, whereas their bulk counterparts possess lower-energy indirect gaps. Many MX2 materials display both metallic and semiconducting phases [55]. The stable step for MX2 at ambient temperature is typically 2H. However, the 1T phase can be induced through Li-intercalation [56][26] or electron beam irradiation [57]. Cmically exfoliated 1T MoS2 exhibits conductivity 107 times higher than the semiconducting 2H phase [58]. Additionally, the 1T or 1T0 phase of WTe2 is more stable than the 2H phase at ambient temperature [59][49]. In the case of MoTe2, the 2H and 1T0 steps can easily transition between each other due to their comparable cohesive energy difference. Moreover, titanium (Ti), chromium (Cr), nickel (Ni), zinc (Zn), vanadium (V), niobium (Nb), and tantalum (Ta) dichalcogenides purely exhibit metallic behaviour [60]

Most MX2 materials exhibit high mobility due to their absence of dangling bonds. However, mobility can be influenced by factors such as substrate choice, metal contacts, and the presence of grain boundaries. At ambient temperature, MoS2 demonstrates mobility of 700 cm2 V1 s1 on a SiO2/Si substrate with scandium (Sc) contacts and a range of 33-151 cm2 V1 s1 on a BN/Si substrate when encapsulated [61], [62]. Like graphene, TMDs possess structural flexibility, robustness, and excellent electrical transport properties. Suspended few-layer MoS2 nanosheets exhibit an extraordinarily high Young's modulus (E) of 0.33 ± 0.07 TPa [63]. Single-layer MoS2 has also been found to possess remarkable in-plane stiffness (180 ± 60 N m1) and Young's modulus of 270 ± 100 GPa [64]. In comparison, monolayer MoS2 surpasses stainless steel (204 GPa) and graphene oxide (207 GPa) in terms of Young's modulus[65], thanks to its defect-free nature, high crystallinity, and absence of stacking faults characteristic of atomically thin TMDs[66].

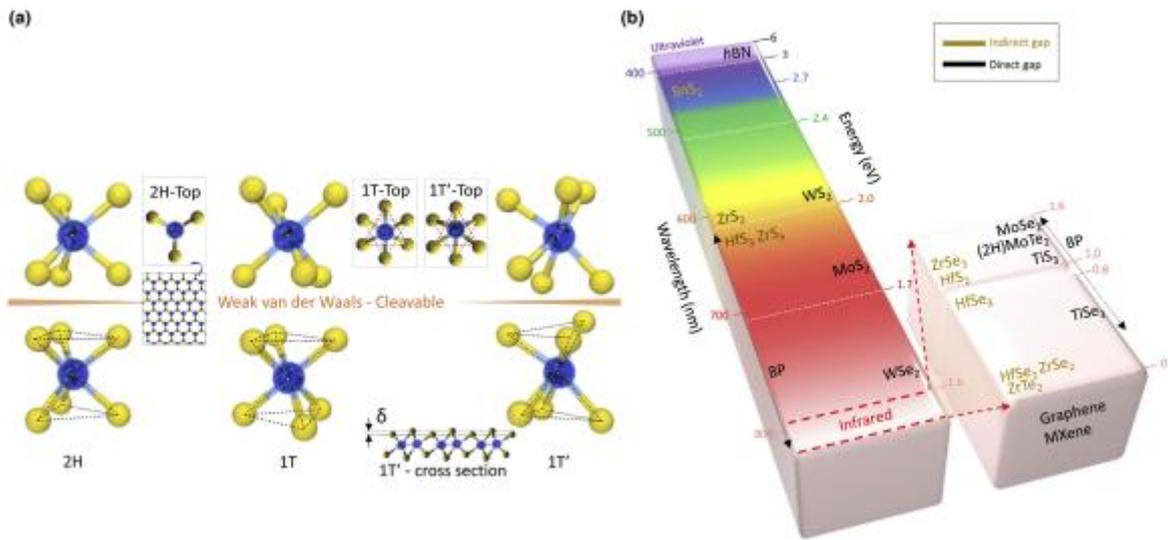

Figure 3 showcases the structures of typical layered transition metal dichalcogenides (TMDs). In multilayer TMDs, cleavable structures such as 2H, 1T, and 1T0 are demonstrated (Fig. 3a). The bandgap of 2D layered materials varies from the zero bandgap of graphene (depicted in white) to the wide bandgap of hBN. The colour in the column represents the wavelength of the bandgap, for instance, the bandgap of MoS2 (1.8 eV) is indicated in red, while WS2 (2.0 eV) is depicted in orange (Fig. 3b). On the left side of the column, indirect materials such as SnS2, ZrS2, HfS3, ZrS3, ZrSe3, HfS2, HfSe3, HfSe2, ZrSe2, and ZrTe2 are displayed. Direct bandgap materials such as h-BN, WS2, MoS2, WSe2, MoSe2, 2H-MoTe2, TiS3, and TiSe3 are presented on the right side.

Recent advancements in synthesising single-component 2D layers have expanded to include group III and V elements. Atomically thin boron layers, known as 'borophenes,' were recently synthesised in an ultrahigh vacuum system by evaporating pure boron elements at high temperatures (450-700°C) on silver (Ag) (111) substrates. Scanning tunnelling spectroscopy confirmed anisotropic metallicity, even though bulk boron allotropes are semiconductors [67]. Bulk black phosphorus (BP) was produced through chemical vapour transport of red phosphorus in the presence of a transport agent [68] or via pressurisation (>1.2 GPa, 200°C) [69], [70]. BP films are excellent candidates for electrical devices due to their high mobility (1000 cm$^2$ V$^{-1}$ S$^{-1}$) and ambipolarity [71]–[73]. The metallic behaviour of MXenes, layered metal carbides/nitrides positioned at the bottom of the schematic, is also highlighted, similar to graphene[74].

Figure 4 provides an overview of the structure and features of 2D TMDs, including charge density wave (CDW), magnetism (ferromagnetic and antiferromagnetic), and superconductivity. Due to the scope of this evaluation, a detailed discussion of each material's specific features is beyond its limits. In addition to TMDs, borophene, silicene, germanene, and stanene are emerging as unconventional 2D materials with a wide range of fascinating characteristics. However, these materials are volatile in ambient air[75], requiring encapsulation or hydrogen termination to form SiH or GeH in silicene or germanene. Borophene exhibits metallic properties, while silicene possesses a 1.9 meV bandgap, germanene has a 33 meV bandgap, and stanene has a 101 meV bandgap, showcasing an inverse trend with an atomic number[76]. Recent research has shown the potential of silicene as a field-effect transistor (FET), making it promising for future electronics [77], [78]. Stanene is considered a strong contender for 2D topological layers.

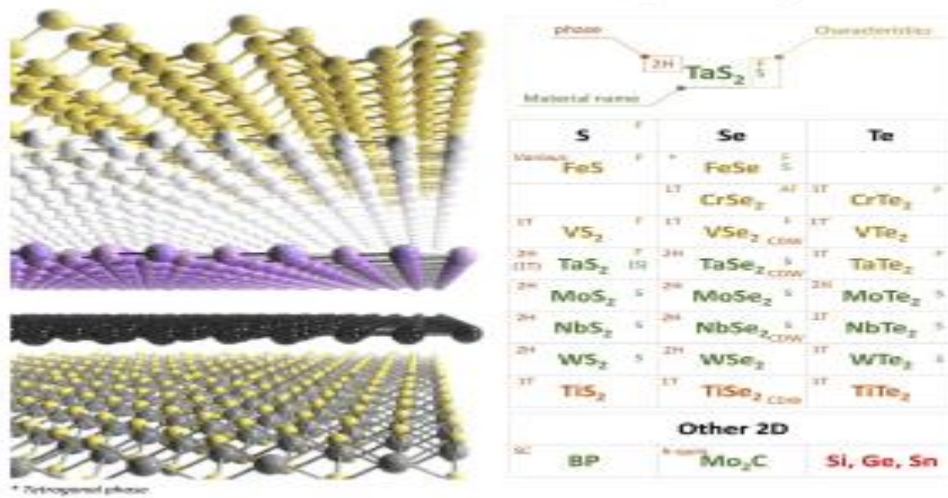

Figure 4 presents a comprehensive table showcasing several 2D TMDs and other 2D materials, highlighting their diverse physical properties, including magnetism (ferromagnetic (F) or antiferromagnetic (AF)), superconductivity (s), charge density wave (CDW), and crystal forms (2H, 1T).

### 3. Recent advances in synthesis processes for 2D TMDs

Considerable effort has been dedicated to synthesising controllable, large-scale, and uniform atomic layers of different 2D TMDs using various top-down and bottom-up techniques, including mechanical exfoliation, chemical exfoliation, and chemical vapour deposition (CVD)[79]. Mechanical exfoliation, renowned for its exceptional quality, has been extensively employed in foundational studies on the fundamental physics and devices of 2D TMDs. However, its progress has been hindered by flake size and film uniformity limitations. Conversely, CVD has emerged as a promising approach for the scalable and reliable production of large-area 2D TMDs[80]. Nonetheless, CVD-produced TMDs tend to exhibit lower quality than their exfoliated counterparts.

Recent attempts have focused on achieving high-quality TMDs with controlled thickness and wafer-scale uniformity through techniques such as atomic layer deposition (ALD), metal-organic chemical vapour deposition (MOCVD), and direct deposition methods (sputtering, pulsed laser deposition, and e-beam)[79]. The synthesis of 2D materials involves incorporating thermal energy from a heated substrate or non-thermal energy, such as microwave or photon energy, into chemical reactions. Factors like substrate lattice parameters, temperatures, and atomic gas flux influence the process. This section will examine the advantages and disadvantages of 2D TMD development techniques, such as CVD, MOCVD, and ALD.

**4. Chemical vapour deposition/vapour phase growth process**

Chemical vapour deposition (CVD) is an efficient technology for producing large-area, atomically thin 2D TMDs, enabling successful device applications. The most common type of CVD employed in developing 2D TMDs involves the co-evaporation of metal oxides with chalcogen precursors, leading to a vapour phase reaction and the formation of stable 2D TMDs on a suitable substrate[81]. The growth mechanism in CVD varies depending on the synthesis process, as several factors influence it: (1) substrate properties, (2) temperature, and (3) atomic gas flow, briefly discussed in the following section. (1) Substrate properties: The nanoscale surface morphology, terminating atomic planes of the substrate, and lattice mismatching play a role in determining the atomic layer arrangement of 2D materials. The substrate's surface energy has been found to impact the nucleation and growth of 2D TMDs. (2) Temperature: The growth temperature critically influences the reaction process. At sufficiently high temperatures, randomly placed adatoms migrate to energetically favourable regions, forming three-dimensional (3D) islands. Conversely, at lower temperatures, the kinetic energy of adatoms is insufficient for effective diffusion, leading to the development of amorphous or polycrystalline films. (3) Atomic gas flux: Achieving high-quality 2D material growth relies on properly controlling the atomic gas flow. Adequate vapour pressure is necessary for mixing atomic vapours and their transport to the substrate[82]. The stability of vaporised atoms is crucial to prevent unwanted reactions during transit. The vaporised atoms are carried to the substrate by a carrier gas, and the flow rate is determined by the Clausius-Clapeyron equation, which relates the rate of change of vapour pressure (P) with temperature (T) to the enthalpy of evaporation (DH) and Boltzmann's constant (k).

Researchers, such as Lee et al. [83], have reported the formation of large-scale MoS2 layers through chemical vapour interaction between molybdenum trioxide (MoO3) and sulfur powder at elevated temperatures (~650°C). The process involves the reduction of MoO3 to generate the suboxide MoO3x, which then combines with vaporised sulfur to produce a 2D layered MoS2 film.

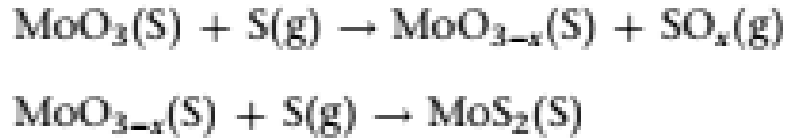

$$MoO_3(S) + S(g) \rightarrow MoO_{3-x}(S) + SO_x(g)$$
$$MoO_{3-x}(S) + S(g) \rightarrow MoS_2(S)$$

This straightforward technique allows for the production of large quantities of MS2, although it often leads to the formation of randomly scattered flakes instead of a continuous film. An interfacial oxide layer has been identified as a significant barrier to the development of MoS2. Similarly, [84] synthesised MoS2 atomic layers on Si/SiO2 substrates through the vapour-phase interaction of MoO3 and S powders, creating triangular MoS2 flakes rather than a continuous layer. The average mobility of these MoS2 flakes was found to be 4.3 cm2 V1 s1, with a maximum current on/off ratio of 106. In Fig. 5a, [85]. Observed an intriguing shape evolution in CVD-grown MoS2 domains, transitioning from triangular to hexagonal geometries depending on the spatial position of the silicon substrate [86] introduced a novel approach utilising MoCl5 and sulfur as precursors to accurately control the number of MoS2 layers across a large area. However, the charge carrier mobility in MoS2-FETs was relatively poor, ranging from 0.003 to 0.03 cm2 V1 s1.

Another technique for producing large continuous areas of TMDs is the "two-step method," which involves depositing a thin film of a transition metal (e.g., Mo, W, Nb) onto a substrate (often Si/SiO2) and then thermally reacting it with chalcogen (S, Se, Te) vapour. This CVD process occurs at high temperatures (300–700°C) in an inert environment, forming stable 2D TMDs through the following reaction.

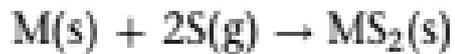

$$M(s) + 2S(g) \rightarrow MS_2(s)$$

This two-step technique has demonstrated the ability to manufacture wafer-sized (2 in.) MoS2 layers and manipulate their thickness on SiO2/Si substrates (Fig. 5b) [87]. The process involves depositing a controlled thickness of metal (W, Mo) onto the substrate and then placing the metal-coated substrate and sulphur powder into a CVD furnace. The reaction atmosphere is maintained inert at 600°C for 90 minutes with a steady flow of 200 ccm Ar. However, monolayer MoS2 produced using this method exhibited point defects and double-layer domains, as confirmed by high-resolution transmission electron microscopy (HRTEM) and Raman analysis. Electrical tests conducted on MoS2 FETs revealed semiconductor behaviour with significantly improved field-effect mobility (12.24 cm2 V1 s1) and current on/off ratio (106) compared to previously reported CVD-grown MoS2-FETs, amorphous silicon (a-Si), or organic thin-film transistors.[88] grew large area MoS2 films using e-beam evaporation and CVD processes and observed p-type conduction, but with extremely low mobilities ranging from 0.004 to 0.04 cm2 V1 s1. It has been demonstrated that including Mo-containing seeds provides a specific region for nucleation and the development of highly crystalline monolayer MoS2 [87]. The presence of seed patterns or molecules on the substrates can control the nucleation of 2D TMDs in predetermined regions, forming large crystals. A MoS2 line island with a diameter of 100 μm was discovered, and the device exhibited carrier mobility and on/off ratio surpassing 10 cm2 V1 s1 and 106, respectively. While the metal-sulfurisation process allows for adjustable thickness and large-scale manufacturing, it is still limited in producing small grain sizes with defects. In addition to the metal film technique, direct sulfurisation/selenization of various metal oxide and chloride precursors such as (NH4)2MoS4, MoO3, WO3, and MoO2 has been commonly employed to generate TMDs. [89] controlled the thickness of WO3 coatings for the large-area production of single-layer and few-layer WS2 sheets. Although (NH4)2MoS4 and related sulfide precursors have been utilised, the control over layer thicknesses has been limited [90].

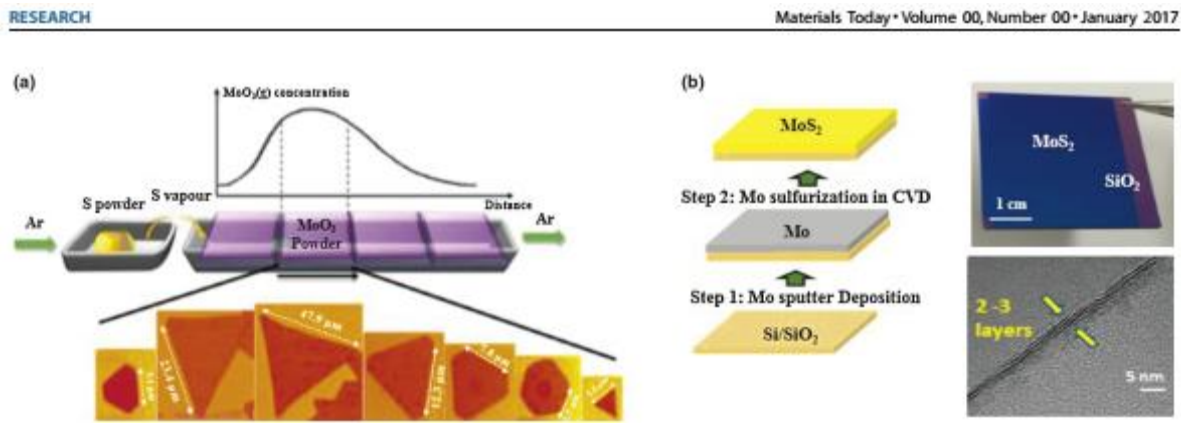

Figure 5 illustrates two important aspects of the research. (a) It presents a schematic diagram of the CVD process and AFM images that showcase the transformation of MoS2 crystals from triangular to hexagonal shapes depending on the spatial position of the silicon substrate. This information is sourced from the publication by Wang et al. [84], [85], 2014, with permission from the American Chemical Society. (b) The figure also displays the successful large-scale production of 2-3 layers of MoS2 in a CVD furnace by utilising a sulfurised Mo seed layer.

This work is referenced from the publication by the American Institute of Physics [86], 2015, with permission.

## 5. Metal-organic chemical vapour deposition (MOCVD)

MOCVD, short for Metal-organic Chemical Vapor Deposition, follows a principle similar to conventional CVD, but with metal-organic or organic chemical precursors [91], [92]. In this technique, complex organic molecules are mixed with the required atoms and transported over a substrate. Upon exposure to heat, the organic molecules decompose, allowing the atoms to be deposited on the substrate one by one. By carefully controlling the atomic composition, high-quality films with desirable crystallinity can be achieved. Figure 6 presents an illustrative schematic of the MOCVD process, outlining the various steps involved in synthesising 2D materials[93].

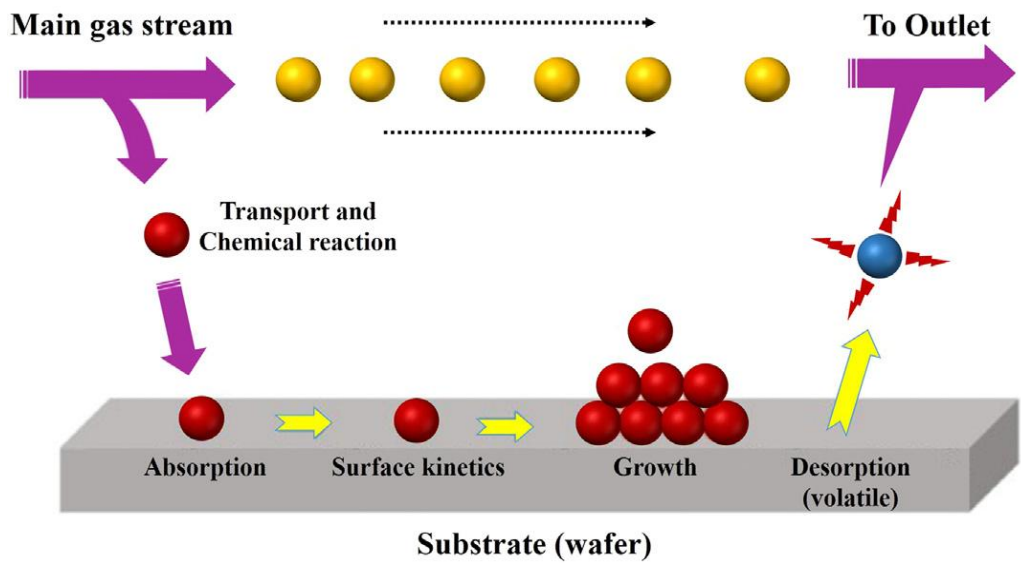

Figure 6 illustrates the MOCVD deposition process, which involves generating active layers on the substrate. The gaseous precursors undergo thermal degradation to produce high-quality thin films and are then deposited onto the substrate. This deposition is followed by surface diffusion kinetics, ensuring the uniformity and proper arrangement of the deposited atoms or molecules on the substrate surface.

MOCVD has emerged as a promising technique for the fabrication of 2D TMDs. It offers several advantages in producing these materials: (i) it enables large-scale and uniform growth of 2D TMDs, and (ii) it provides precise control over both metal and chalcogen precursors, allowing manipulation of the composition and morphology of 2D TMDs. [94]. employed molybdenum hexacarbonyl (Mo(CO)6), tungsten hexacarbonyl (W(CO)6), ethylene disulfide ((C2H5)2S), and H2 gas-phase precursors in combination with an Ar gas carrier to fabricate monolayer and few-layer MoS2 and WS2 films on SiO2 substrates, covering wafer-scale areas (4-inches). The team successfully demonstrated the production of large-scale MoS2 and WS2 films on 4-inch fused silica substrates (Fig. 7a). It manufactured approximately 8000 MoS2 FET devices using standard photolithography techniques (Fig. 7b). The MoS2-FETs exhibited electron mobility of 30 cm2 V1 s1 at room temperature and 114 cm2 V1 s1 at 90 K (Fig. 7c). The time evolution of monolayer coverage across the entire substrate as a function of critical time is depicted in Figure 7d (t0).

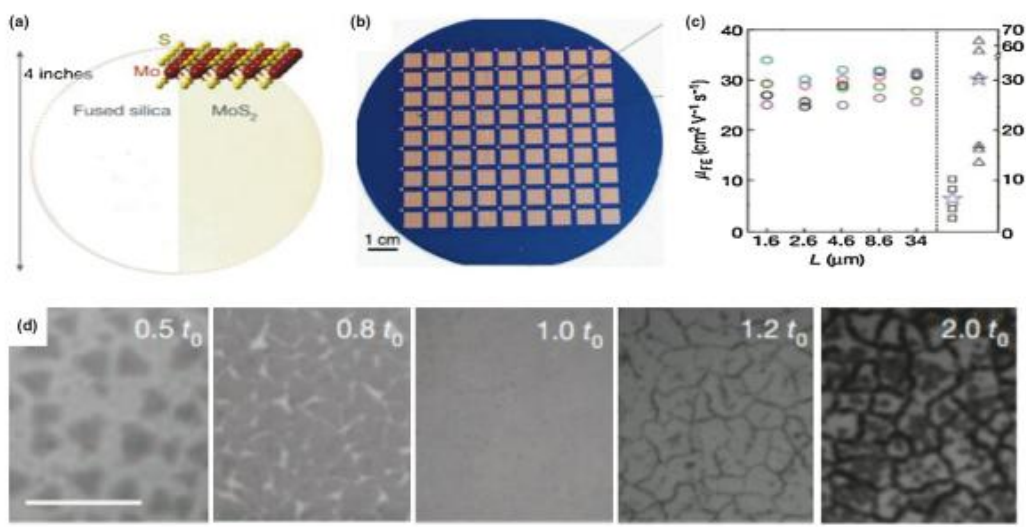

Figure 7 showcases the results of the study. (a) Large-scale growth of continuous MoS2 monolayers on fused silica was achieved using all-gas phase precursors. (b) The scalability of the process enabled the mass production of 8000 field-effect transistor (FET) devices. (c) Field-effect mobility (mFE) was measured from FET devices of various lengths, showing a consistent mobility of 30 cm2 V1 s1. (d) Optical images of MoS2 films captured at different growth times illustrate the evolution of coverage, with t0 representing the optimal growth time for complete monolayer formation (scale bar: 10 mm)[93].

Eichfeld et al. [93]. recently reported a significant breakthrough in the large-scale production of mono and few-layer WSe2 using MOCVD with W(CO)6 and dimethyl-selenium ((CH3)2Se) precursors. They demonstrated that various factors, such as temperature, pressure, Se: W ratio, and substrate choice, played a crucial role in determining the morphology of WSe2 films. Figure 8a illustrates the diverse morphologies of WSe2 on different substrates, including epitaxial graphene, CVD graphene, sapphire, and BN. Notably, WSe2 exhibited high nucleation density on graphene, while sapphire showed the largest domain size, ranging from 5 to 8 mm. The influence of gas flow rate on the size and shape of WSe2 domains on epitaxial graphene is depicted in Figure 8b. In addition to the flow rate, the domain size increased with higher pressure and temperature. For instance, as the temperature increased from 800 to 9008C, the grain size grew by 200 per cent, from 700 nm to 1.5 mm. Similar trends were observed on sapphire substrates[95]. The Se: W ratio and total gas flow through the system also significantly impacted the domain size of WSe2 films. By increasing the Se: W ratio from 100 to 2000, the domain size could be increased from 1 to 5 mm.

Similarly, increasing the total gas flow from 100 to 250 ccm resulted in domain sizes of up to 8 mm[96]. The I-V characteristics confirmed the formation of a tunnel barrier for vertical transport created by WSe2, indicating the presence of a clean van der Waals gap in WSe2/graphene heterostructures. Furthermore, the MOCVD technique was utilised to fabricate MoS2/WSe2/graphene and WSe2/MoS2/graphene heterostructures (Fig. 8c) [97]. Surprisingly, directly produced heterostructures by MOCVD exhibited resonant tunnelling of charge carriers, leading to high negative differential resistance (NDR) at room temperature, as shown in Fig. 8d. Although the MOCVD process offers flexibility, scalability, and precise control over film stoichiometry, the use of hazardous precursors, slow film growth rate, and high production costs limits its broader application.

6. **Atomic layer deposition (ALD)**

Atomic Layer Deposition (ALD) is a gas-phase chemical technique that enables the precise fabrication of atomically thin films by sequentially reacting precursors with the substrate. While ALD has been predominantly used for oxide materials, several research groups have successfully explored binary sulfide materials, including $TiS_2$, $WS_2$, $MoS_2$, SnS, and $Li_2S$. This study focuses on the most significant outcomes of the ALD approach for 2D TMDs.[98]demonstrated excellent control over the thickness of $MoS_2$ films by utilising self-limiting reactions of molybdenum pentachloride ($MoCl_5$) and hydrogen sulfide ($H_2S$) on sapphire substrates. However, high-temperature annealing at 800°C was required to achieve large triangular $MoS_2$ crystals measuring 2 mm.[99] Wafer-scale production of $WS_2$ was achieved through ALD of tungsten oxide ($WO_3$), followed by conversion through $H_2S$ annealing. Figure 9a illustrates the ALD growth processes for $WS_2$ nanosheet fabrication, and the number of $MoS_2$ layers can be precisely controlled by adjusting the number of ALD cycles for $MoO_3$ deposition. In Figure 9b, a camera image showcases large-area mono-, bi-, and tetralayer $WS_2$ nanosheets on $SiO_2$ substrate, spanning approximately 13 cm long. Top-gate monolayer $WS_2$ FETs exhibited n-type conduction with electron mobility of 3.9 cm2 V1 s1 (Fig. 9c). Furthermore, leveraging ALD's excellent conformal growth capabilities, they successfully achieved the controlled development of 1D $WS_2$ nanotubes (WNTs) by sulfurising $WO_3$ layers on Si nanowires (NWs) [64] introduced another chemical approach for depositing $MoS_2$ on $SiO_2$/Si substrates using $Mo(CO)_6$ and dimethyldisulfide ($CH_3SSCH_3$, DMDS) as Mo and S precursors, respectively, following the first report on ALD synthesis of 2D materials. The as-deposited samples were annealed at 9008°C to crystallise into the 2H-

MoS2 phase. These preceding experiments indicate that the growth temperature was relatively high (800–1000°C), and the resulting films exhibited crystallite sizes in the sub-10 nm range. Consequently, developing an advanced ALD technique for producing high-quality, large-scale MX2 materials with precise control over atomic layer thickness is of paramount importance [100] achieved low-temperature (300–4508°C) synthesis of WS2 atomic layers using WF6 and H2S precursors, along with Si and H2 plasma reducing agents for CVD and ALD, respectively[64]. These layers were formed without template layers or post-deposition annealing treatments [101]. ALD offers scalability and precise thickness control at low substrate temperatures. However, challenges such as high cost and the use of sensitive precursors need to be addressed.

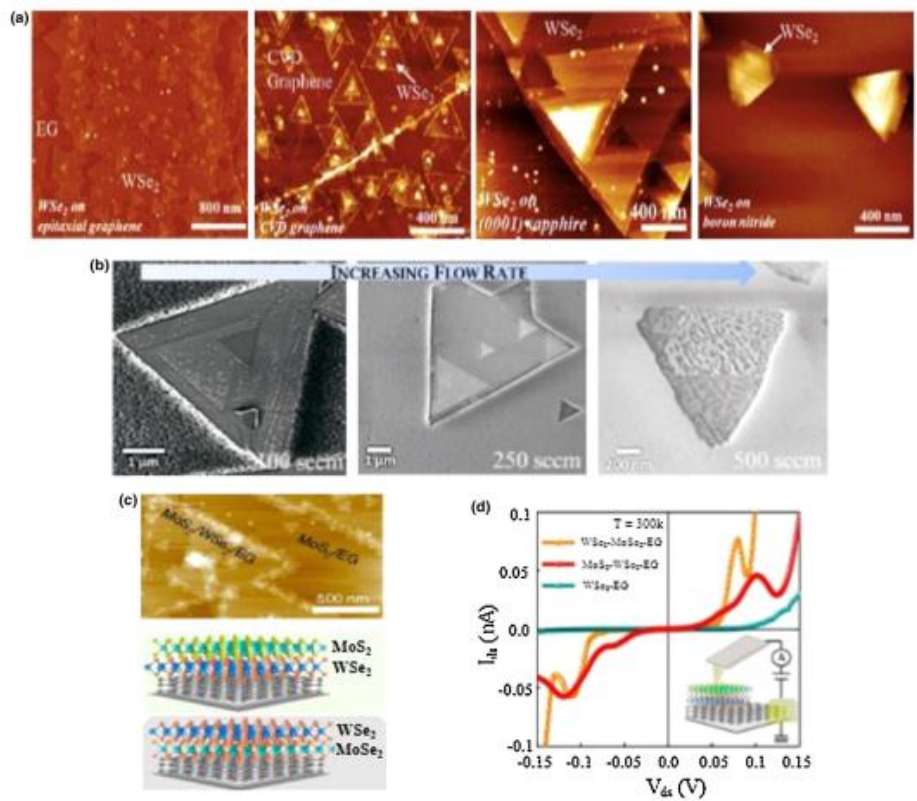

Figure 8 showcases various aspects of WSe2 growth and heterostructure formation. In (a), the AFM surface morphology of WSe2 sheets grown on different substrates, including epitaxial graphene, CVD graphene, sapphire, and boron nitride, is depicted. The FESEM images in (b) demonstrate how the domain size of WSe2 increases with the flow rate of the precursors. This figure is reproduced from the American Chemical Society publication [97][94]. Moving to (c), schematic representations and AFM images illustrate different heterostructures formed with WSe2. Lastly, (d) presents I-V curves for several dichalcogenide-graphene heterostructures, revealing the presence of resonant tunnelling and negative differential resistance (NDR). Permission to reprint this figure was obtained from the American Chemical Society [97] in 2013.

## 7. Potential applications

2D transition metal dichalcogenide (TMD) materials have garnered significant interest due to their wide applications in electronics, photonics, sensors, and energy devices. These applications are driven by the unique properties of layered materials, including their ultrathin atomic profile, which enables maximum electrostatic efficiency, mechanical strength, tunable electrical structure, optical transparency, and high sensor sensitivity [102]. Flexible nanotechnology holds great promise for various applications as it explores the potential of 2D materials to revolutionise ubiquitous electronics and energy devices. Flexible technology encompasses diverse scalable and large-area devices, such as thin-film transistors (TFTs), displays, sensors, transducers, solar cells, and energy storage devices, all designed on mechanically compliant substrates.

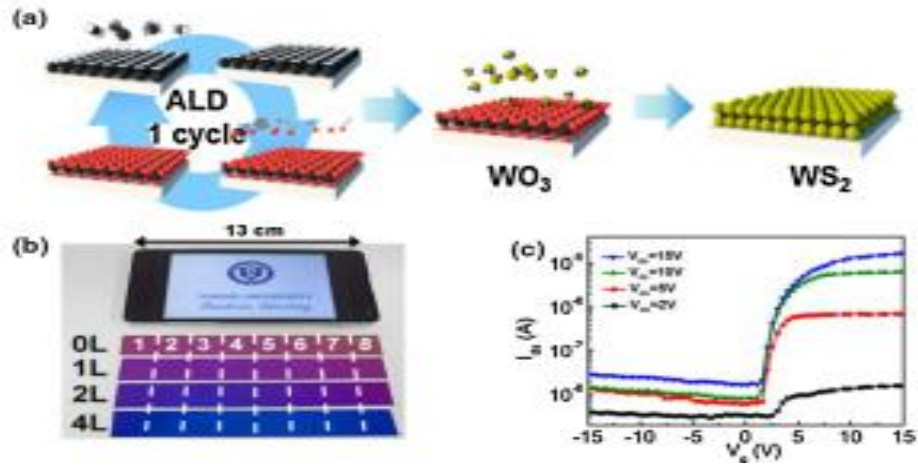

Figure 9. (a) Schematic illustration of the ALD process for synthesising large-area and thickness-controlled WS2 films. (b) Display of the successful synthesis of mono-, bi-, and tetralayer WS2 films over a large area (approximately 13 cm) on Si/SiO2 substrates, with the growing region comparable to the size of a cell phone display screen. (c) Transfer characteristics of a single-layer WS2 field-effect transistor (FET) demonstrating n-type conduction[99].

## 7.1 Electrical and optoelectronic applications:

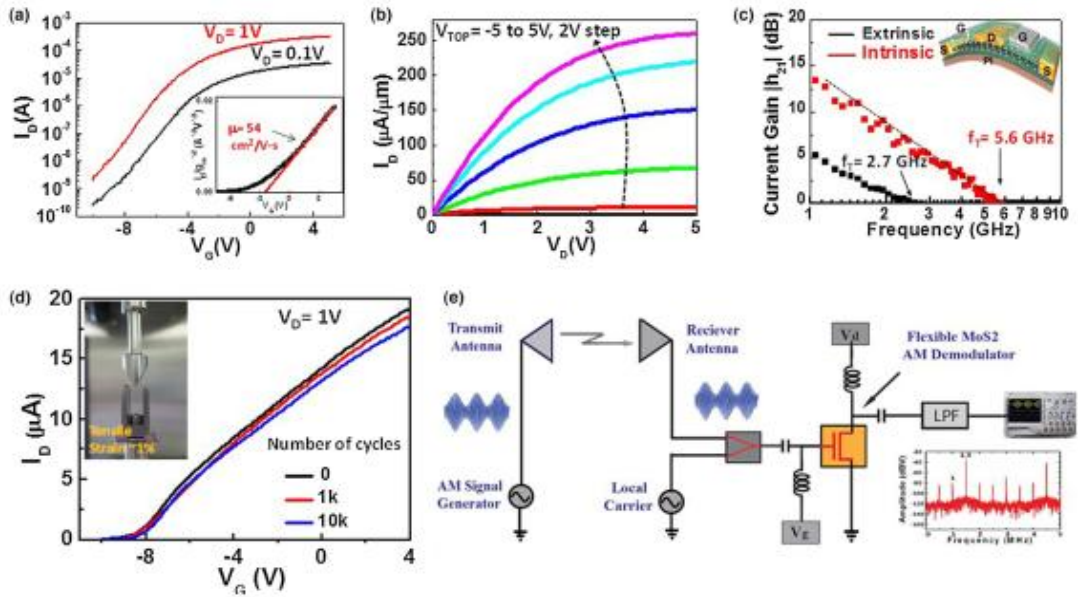

Figure 10. MoS2 field-effect transistors (FETs) are grown by chemical vapour deposition (CVD) on SiO2/Si with a thickness of 280 nm. (a) Electrical transfer properties. The inset

highlights the high low-field mobility of 54 cm2 V1 s1 at the upper end of the range for monolayer MoS2. (b) Output characteristics (ID–VD) showing a linear saturation profile characteristic of well-behaved semiconducting FETs. (c) Calculated cut-off frequency for flexible MoS2 transistors with an inherent fT of 5.6 GHz (L = 0.5 mm). The device structure is depicted in the inset. (d) Bending experiments demonstrate the high electrical stability of flexible MoS2, with 10,000 bending cycles at 1% tensile strain. (e) Circuit diagram of a flexible MoS2 RF transistor used as an AM demodulator in a wireless AM receiver system operating in the AM radio band of 0.54–1.6 MHz. The spectrum of the AM receiver output is shown in the inset. (Reprinted with permission from Ref.[103] , 2015, John Wiley and Sons.)

The limitations of traditional silicon-based technology in terms of scaling have led to the exploration of atomically thin semiconductors like TMDs for future large-scale devices [104]. Over the past decade, significant progress has been made in the research and development of graphene and other 2D materials. Graphene touch panel displays were integrated into smartphones in China as early as 2014, demonstrating the rapid pace of innovation in this field. Several studies have analysed the advancements and potential applications of 2D materials during this period [103], [105]. This section will focus on the advances in flexible electronics, specifically 2D TFTs, which serve as crucial devices for various flexible technology designs similar to their conventional counterparts. High-performance TFTs based on synthesised MoS2 have been successfully developed after years of dedicated research and development. These TFTs exhibit a high on/off current ratio and current saturation and operate at room temperature, showcasing the excellent properties of high-quality TMDs (Fig. 10a,b). Notably, the electron mobility of 50 cm2 V1 s1 and current density of 250 mA/mm have been achieved, which are highly promising for high-performance TFTs. Surprisingly, cut-off frequencies greater than 5 GHz have been attained on flexible plastic substrates with a channel length of 0.5 millimetres (Fig. 10c). Although MoS2 has relatively low mobility, at the high electric fields required for maximum high-frequency operation, transport is determined by the saturation velocity (v-sat), which is sufficiently reasonable (2 × 106 cm/s) to achieve GHz speeds at sub-micron channel lengths [106].

Additionally, flexible monolayer MoS2 TFTs have demonstrated reliable electrical performance even after thousands of mechanical bending cycles (Fig. 10d). These features make MoS2 and similar TMDs highly appealing for low-power RF TFTs in advanced flexible Internet of Things (IoT) and wearable nano-systems, owing to their high on/off ratio, saturation velocity, and mechanical strength. As a result, these materials hold significant potential for developing low-power RF TFTs for improved flexible IoT and wearable devices. An example is a wireless flexible radio system where monolayer CVD MoS2 is employed to demodulate the received signal (Fig. 10e). Emerging 2D materials, like BP, hold immense promise for achieving higher speeds and frequencies in transistor operation. These materials have demonstrated remarkable transistor mobility, reaching up to 1000 cm2 V1 s1 (Fig. 11a), comparable to the mobility of CVD graphene. Additionally, they exhibit on/off ratios typically exceeding 102–103 at room temperature [107].

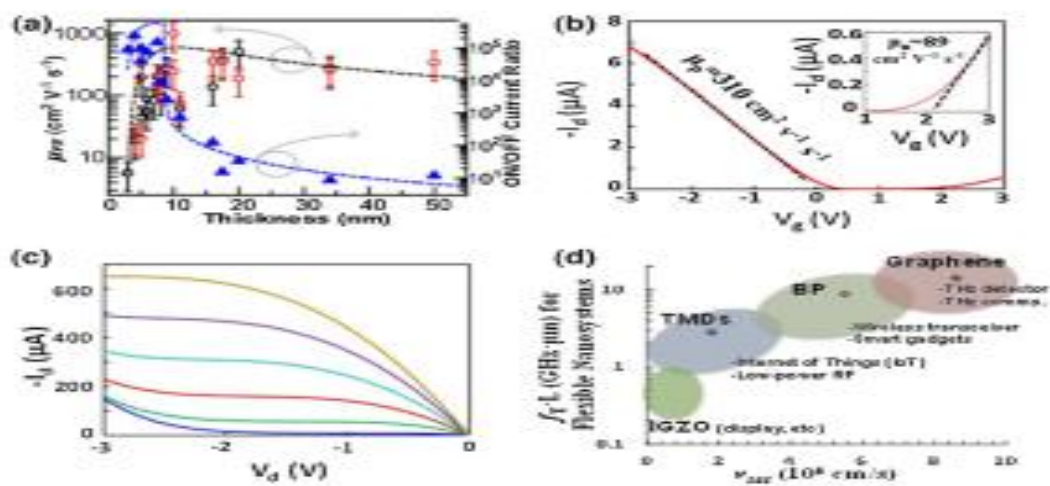

Figure 11 showcases several important aspects of BP and its potential applications. (a) It demonstrates the mobility due to field-effect (mFE) with open circles and the Ion/off ratio of BP films of varying thicknesses with filled blue triangles [107]. (b) Transfer characteristics of an encapsulated BP ambipolar TFT on polyimide are shown, revealing a field hole mobility of 310 cm2 V1 s1, the electron mobility of 89 cm2 V1 s1, and an on/off ratio exceeding 103. The device parameters include Vd = 10 mV, flake thickness of 15 nm, and W/L ratio of 10.6 mm/2.7 mm. (c) Current saturation output curves from the same device are displayed, illustrating gate biases ranging from 0 to 2.5 V from bottom to top [106]. (d) Additionally, examples of adaptable high-frequency nano-system applications are presented based on experimentally obtained saturation velocities, incorporating modern organic and metal-oxide materials [106].

BP, a unique 2D crystal with intermediate optoelectronic properties between high-mobility graphene and low-mobility TMDs, holds significant promise for high-performance flexible nano-optoelectronics. In 2015, the first flexible BP TFTs were introduced, exhibiting ambipolar transport and surpassing the hole and electron mobilities of thin-film materials based on metal oxides, organic semiconductors, and amorphous Si (Fig. 11b) [107]. Remarkably, these flexible BP TFTs were not optimised regarding contacts and interfaces, suggesting even more significant potential for performance enhancement. The output characteristics of these TFTs demonstrate excellent current saturation (Fig. 11c). Leveraging this feature, flexible BP TFTs have been utilised in the construction of inverting and non-inverting analogue amplifiers, digital inverters, and frequency multipliers on flexible substrates [106], [107]. Figure 11d illustrates the high-frequency application domain for flexible nano-systems based on the experimental saturation velocity of various 2D materials, ranging from TMDs to graphene. TMDs are well-suited for low-power RF and IoT systems, while BP, with its higher vsat, shows promise for wireless microwave and 5G systems. Lastly, with its exceptional mobility and vsat, graphene enables the realisation of flexible THz detectors and communication systems [106], [108].

**7.2 Energy applications**

Due to their unique atomically layered structure, large surface area, and exceptional electrochemical properties, two-dimensional transition metal dichalcogenides (2D TMDs) are increasingly recognised as promising electrode materials for energy storage devices such as supercapacitors and Li-ion batteries. The multilayer architecture of these materials provides additional ion storage sites while maintaining structural stability during charge and discharge cycles. Combined with their surface functionality and electrical conductivity, the high surface area of 2D materials, such as graphene, with its surface area of 2630 m2/g (the highest among carbon materials), makes them highly suitable for energy storage electrodes [89], [109], [110].

Supercapacitors, electrochemical capacitors with capacitance values one order of magnitude higher than Li-ion batteries consist of two symmetric electrodes separated by a membrane and an electrolyte that facilitates ionic connectivity between the electrodes. When voltage is applied, ions in the electrolyte form electric double layers with polarity opposite to the electrode's charge. In some electrode materials, specific ions can penetrate the double layer and undergo selective adsorption, contributing to the overall capacitance with pseudo-capacitance. $MoS_2$, with its stacked-sheet-like structure and a wide range of Mo oxidation states (+2 to +6), exhibits high electrical double-layer capacitance (EDLC) as well as significant pseudo-capacitance, making it a promising candidate for supercapacitor electrodes in the realm of 2D materials [111]–[113]. However, challenges such as small flake size, limited yield, unpredictable thickness, and defects arising from exfoliation and hydrothermal processes need to be addressed for its widespread implementation [114], [115].

The electrochemical performance of MoS2 nanosheets can be enhanced by controlling their surface morphology. Tour et al.[114]. engineered edge-oriented/vertically aligned MoS2 nanosheets with more extensive van der Waals gaps and reactive dangling bond sites, creating favourable conditions for electrolyte ions and resulting in high capacitive characteristics (Fig. 12a). These vertically aligned MoS2 nanosheets resembling a sponge-like structure, exhibit remarkable areal capacitance of up to 12.5 mF cm2. It has been acknowledged [115], [116]. that the low electrical conductivity of the prevalent 2H-MoS2 phase limits its suitability as a supercapacitor electrode material. To address this limitation, researchers at Rutgers University developed a metallic phase (1T) of MoS2 with 107 times higher conductivity than the semiconducting phase (Fig. 12b). Super capacitors based on chemically exfoliated MoS2 nanosheets as electrodes demonstrated exceptional capacitive performance across various aqueous electrolytes, achieving capacitance values ranging from 400 to 700 F cm3 [117].

In a different approach, Choudhary et al. [118] employed magnetron sputtering techniques to directly fabricate a large-scale and unique 3D-porous MoS2 supercapacitor electrode on flexible substrates such as copper and polyimide (Fig. 12c). The distinctive 3D porous structure facilitated significant intercalation of electrolyte ions, resulting in a high areal capacitance of 33 mF cm2 (equivalent to 330 F cm3) at a discharge current density of 25.47 mA/cm2. Furthermore, the electrode demonstrated exceptional cyclic stability, retaining 97% of its capacitance after 5000 cycles. On the other hand, MXenes exhibit outstanding electrical conductivity and hydrophilicity, enabling them to achieve a substantial volumetric capacitance of 900 F cm3 [60], surpassing the performance of MoS2 in this aspect.

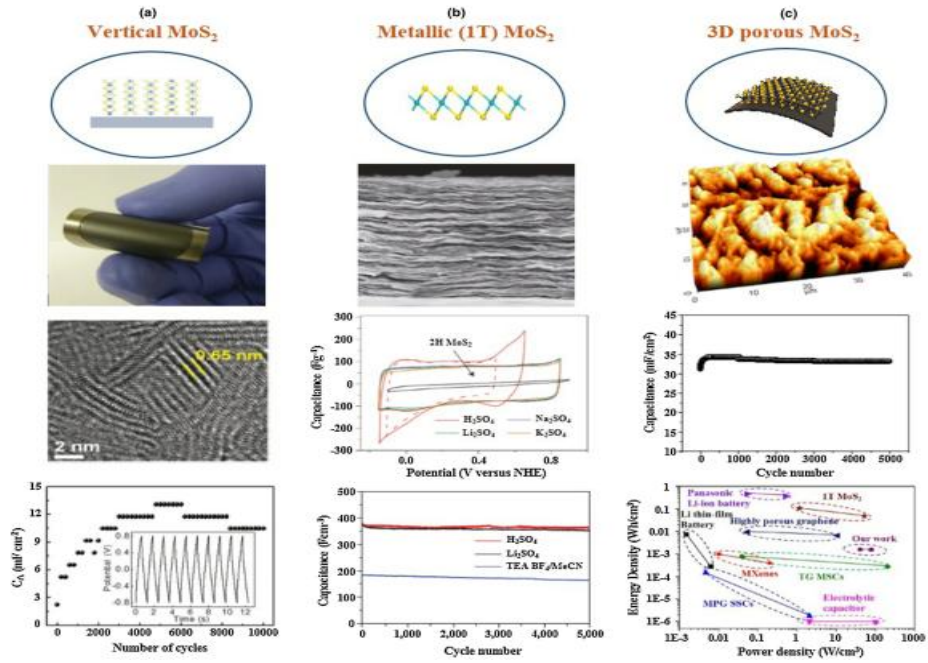

Figure 12 showcases significant developments in MoS2 electrode materials for energy storage applications. In (a), a flexible MoS2 electrode is depicted along with a transmission electron microscopy (TEM) image of the edge-oriented MoS2 film. The cyclic stability curve demonstrates the impressive capacitance retention of the electrode over 10,000 cycles [119]. Moving to (b), a scanning electron microscopy (SEM) image displays the chemically exfoliated 1T MoS2 electrode. The electrochemical performance of the 1T and 2H phases is compared through cyclic voltammetry (CV) results, and the cyclic stability of the 1T-MoS2 electrode is demonstrated[120]. Finally, (c) showcases the large-scale manufacturing of MoS2 on flexible substrates, with the atomic force microscopy (AFM) image revealing the 3D-porous structure of the MoS2 material[31].

Existing Li-ion batteries, which rely on graphite anodes and lithium cobalt oxide (LiCoO2), have specific capacities that fall short of current demands, with theoretical and observed capacities of 372 mAh/g and 150 mAh/g, respectively [121], [122]. To address this limitation, 2D MoS2 has emerged as a promising anode material for Li-ion batteries. It offers a high theoretical capacity of 670 mAh/g and possesses large van der Waals gaps between its layers, enabling the intercalation of Li+ ions without significant volume changes or active material disintegration during charge and discharge processes [123], [124]. The electrochemical reaction of Li with MoS2, represented by the equation MoS2 + 4Li+ + 4e → 2Li2S + Mo, allows for the storage of twice the capacity compared to graphite electrodes due to the involvement of four moles of Li per mole of MoS2. However, despite the attractive specific capacity exhibited by various MoS2 structures, concerns persist regarding their poor cycle stability and low electronic conductivity [125], [126]. Composite designs combining MoS2 with highly conductive nanomaterials like graphene and carbon nanotubes (CNTs) have been explored to address these issues. For instance, Wang et al. developed an exfoliated MoS2-C composite with a capacity of 400 mAh/g at 0.25 C [127]. Additionally, Patel et al. successfully merged 3D CNTs with MoS2 to create a hybrid anode material with vertical flake morphology, resulting in excellent electrochemical performance, including an aerial capacity of 1.65 mAh/cm2 (450 mAh/g) at a 0.5 C rate and cycling stability up to 50 cycles at 0.5 C[128].

In the context of sodium-ion batteries (SIBs), which have been proposed as an alternative to LIBs, TMDs like MoS2 and WS2 are considered promising host anode materials due to their layered architectures that can accommodate the larger size of Na+ ions [129]. Moreover, MXenes have shown great potential as efficient electrode materials in Li-ion batteries. For example, Mashtalir et al. demonstrated a capacity of 410 mAh/g at a 1 C rate using delaminated Ti3C2-based MXene [130].

## 8. 2D TMDs sensors

The demand for susceptible, selective, low-power, reliable, and portable sensors has fueled extensive research on novel sensing materials based on 2D allo-electronics. Transition metal dichalcogenides (TMDs) and phosphorous-based materials have gained popularity following the success of their 2D carbon counterpart, graphene[131]. The large surface-to-volume ratio in 2D TMDs offers tremendous potential for detecting many target analytes per unit area, enabling fast response and recovery with minimal power consumption [132], [133]. Moreover, the recent advancements in the scalable synthesis of 2D TMDs have demonstrated the feasibility of producing cost-effective sensors. Figure 13 illustrates the utilisation of 2D TMDs such as MoS2, WS2, and others in various sensing applications, including gas, chemical, and biosensors.

As expected, most of the reported sensors based on 2D TMDs [134], [135]have been developed using mechanically or liquid-phase exfoliated MoS2 flakes. For instance, [32] fabricated a MoS2-FET sensor device to detect nitric oxide (NO) using mechanically exfoliated single and few-layer MoS2 films. The MoS2 films exhibited high sensitivity to NO, with a detection limit of 0.8 parts per million (ppm). Donarelli et al.[135] demonstrated that a sensor based on chemically exfoliated MoS2 flakes outperformed similar sensors, achieving a measured detection limit of 20 parts per billion (ppb) when exposed to nitrogen dioxide (NO2) gas. The p-type behaviour of MoS2 towards NO2 can be attributed to N substitutional doping in S vacancies on the MoS2 surface. This detection limit for NO2 is the lowest reported for MoS2-based sensors and is comparable to or better than sensors based on materials such as ZnO, graphene oxide, and carbon nanotubes. Black phosphorus (BP) and its single atomic layer, phosphorene, have also shown great potential for gas detection, on par with other 2D materials like graphene and MoS2[136], [137]. Hui et al. [33]observed a detection sensitivity of up to 20 ppb in their mechanically exfoliated phosphorene-based sensor.

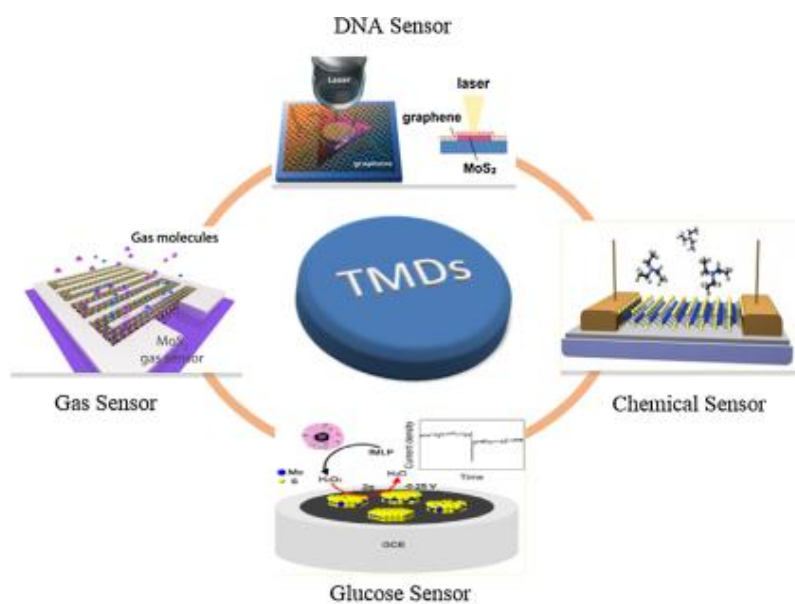

Figure 13 showcases a range of gas, chemical, and biosensors that have been developed using 2D TMDs materials, including MoS2, WS2, and others. These sensors demonstrate the versatility and potential of TMDs in sensing applications. The figure includes illustrations from various sources, including Ref. [138] and Ref. [139] from the American Chemical Society, Ref. [140] from John Wiley & Sons, and Ref. [141] from the Nature Publishing Group, all used with permission.

Abbas et al. [142] conducted a study using a nanosheet (PNS)--based FET device to detect NO2 in a dry air environment. Their research demonstrated impressive sensitivity down to 5 ppb by employing FETs based on few-layer phosphorene. In Figures 14a and b, the schematic of the multilayer phosphorene FET sensor and its sensing performance at various NO2 concentrations are illustrated. The phosphorene sensor exhibited a remarkable response to NO2 concentrations as low as 5 ppb, resulting in a 2.9% change in conductance. Another notable study by Perkins et al. [138] involved the utilisation of single-layer MoS2 as chemical sensors. They developed planar FET sensors using monolayer MoS2 on Si/SiO2 wafers, which displayed excellent responsiveness when exposed to a diverse range of analytes, including laboratory chemicals, nerve gas agents, and various solvents like dichlorobenzene, dichloropentane, nitromethane, nitrotoluene, and water vapour.

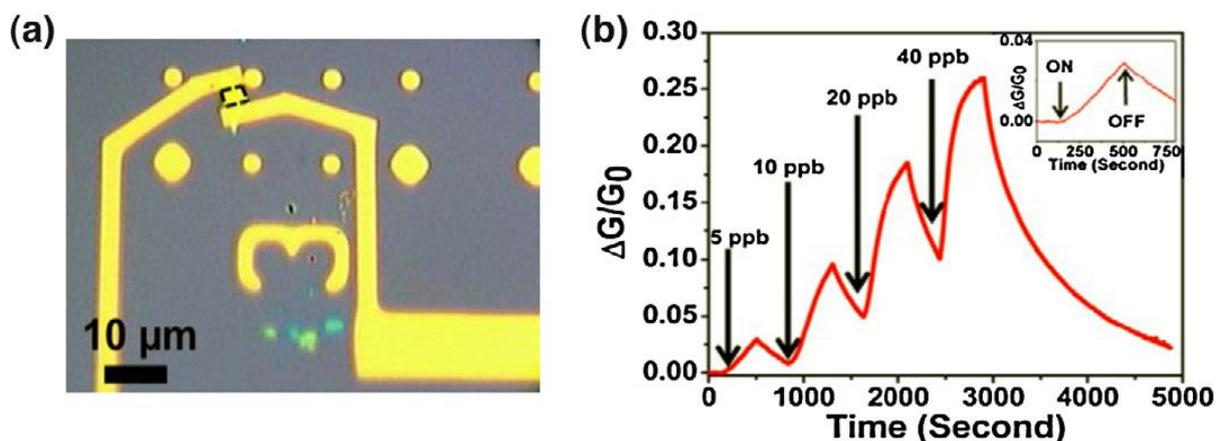

Figure 14 depicts (a) a photograph of multilayer phosphorene FETs featuring Ti/Au electrodes used for chemical sensing. A dashed black line highlights the thin film flake. (b) The gas sensing performance of multilayer phosphorene for NO2, represented by the relative conductance change (ΔG/G0) over time in seconds for different NO2 concentrations ranging from 5 to 40 ppb. The inset provides a magnified view of the response to 5 ppb NO2 exposure, indicating when the NO2 gas is turned on and off. This image is reproduced with permission from the American Chemical Society [142], 2015.

Biosensing is another significant area where 2D transition metal dichalcogenides (TMDs) have experienced remarkable growth in recent years. Due to their larger surface area, 2D TMDs offer the advantage of immobilising a more significant number of biomolecules per unit area, enabling highly efficient biosensors capable of detecting various biomolecules like DNA, glucose, dopamine, and hydrogen peroxide, among others[139][143]–[146] Compared to 3D bulk materials and 1D nanomaterials such as silicon nanowires or carbon nanotubes, 2D TMDs exhibit superior sensitivity, device scalability, flexibility, and transparency in biosensing applications. For instance, Sarkar et al. [35] developed label-free FET-based biosensors using MoS2, demonstrating pH detection sensitivity at least 74 times higher than graphene-based biosensors. Numerous studies have also reported the successful detection of DNA.

MoS2-based biosensors[145]–[147].

| Material | Fabrication method | Gas/chemical tested | Sensitivity | Operating temperature (°C) | Response time |
|---|---|---|---|---|---|
| **Semiconductor metal oxides** | | | | | |
| SnO$_2$ | Hydrothermal [157] | Ethanol | 45.1 ppm | 200 | Minutes |
| ZnO | Vapor-phase transport [158] | H$_2$S | 17.3 ppm | 30 | |
| TiO$_2$ | Electrospinning [159] | CO | 21 ppm | 300 | |
| **Carbonaceous materials** | | | | | |
| CNTs | SWCNT dispersion [160] | NH$_3$ | 3 ppb | R.T. | 1–5 s |
| Graphene | CVD [161] | NO and NO in N$_2$ | 158 ppb | R.T. | |
| **MoS$_2$** | Chemical exfoliation [142] | NO$_2$ | 20 ppb | R.T. | (5–30 s) |
| **Phosphorene** | Mechanical exfoliation [145] | NO$_2$ | 5 ppb | R.T. | |

Table. 1. We comprehensively compare gas/chemical and biosensors based on transition metal dichalcogenides (TMDs) with conventional sensors utilising metal oxides and carbonaceous materials. The comparison considers various parameters such as sensitivity, selectivity, response time, device scalability, and cost-effectiveness. By evaluating these factors, we can assess the advantages and potential of TMDs in sensor applications compared to traditional sensor materials.

Farimani et al. [148]. demonstrated the exceptional capability of monolayer MoS2 nanopores in detecting DNA bases. Compared to graphene nanopores, MoS2 nanopores exhibited a more distinct signal per base and a significantly lower signal-to-noise ratio, making them highly effective for DNA detection. However, their vulnerability to moisture and oxygen hinders the long-term use of MoS2-based biosensors, leading to a decrease in their functionality. To address this issue, Chen and colleagues [155] developed graphene/MoS2 heterostructures, where graphene acted as a protective barrier between MoS2 and the surrounding environment while enhancing the detection of foreign molecules as a host. Similarly, WS2 has recently been found to possess biosensing capabilities for single-stranded DNA (ssDNA) chains through fluorescence quenching [140].

2D TMDs, such as MoS2 and phosphorene, exhibit remarkable sensitivity of up to 20 ppb and 5 ppb, respectively, with rapid response times ranging from 5 to 30 seconds. They outperform most conventional metal oxide-based sensors, which suffer from extended recovery times and high operating temperatures [148]–[150]. Moreover, the performance and sensitivity of MoS2-based FET biosensors have been reported to be 74 times higher than state-of-the-art graphene-based biosensors. To compete with carbonaceous materials like carbon nanotubes (CNTs) and graphene, the gas and chemical sensing capabilities of 2D TMDs need to be further enhanced [151]. revealed that pure graphene can detect gas molecules at deficient concentrations, with detection limits as low as 158 parts per trillion (ppq), exhibiting high selectivity. A comparison table of 2D TMDs and phosphorene gas/chemical sensors with conventional and carbon-based gas sensors is provided in Table 1.

Despite significant advancements, sensors based on mechanical and liquid exfoliation methods are limited to small-scale production. They are unable to meet the growing demand for industrial-scale manufacturing. Proposed a more scalable approach, demonstrating wafer-scale fabrication of MoS2 sensors prepared via chemical vapour deposition (CVD) with comparable NO2 and ammonia (NH3) sensitivity to exfoliated MoS2-based sensors [152]. In another study [153], the 2D MoS2/graphene heterostructure gas sensor exhibited a detection limit of approximately 1.2 ppm for NO2.

## 9. Conclusion and prospects

This article comprehensively assesses the latest advancements in the production, characterisation, and applications of 2D transition metal dichalcogenides (TMDs). The progress made in 2D TMDs offers exciting opportunities for exploring new phenomena, ideas, and technologies in electronics, optoelectronics, and energy. Semiconducting 2D TMDs have great potential for developing ultra-small, low-power transistors that outperform silicon-based FETs. Moreover, the flexibility of 2D TMDs, demonstrated by their compatibility with flexible substrates, has captured researchers' interest in flexible device applications.

The unique atomically layered structure, large surface area, and excellent electrochemical properties of 2D TMDs make them promising candidates for high-efficiency energy storage, especially when combined with surface functionality and electrical conductivity. Furthermore, the high surface-to-volume ratio of 2D TMDs-based sensors leads to improved sensitivity, selectivity, and power consumption. Additionally, the significant spin-orbit interactions and absence of inversion symmetry at the maximum atomic thickness of TMDs have led to the emergence of a new field called "Valleytronics."

Introducing new 2D material families like black phosphorus (BP), silicene, and germanene holds excellent potential for developing low-dimensional electronic and energy devices, provided that their stability concerns can be addressed. Despite significant progress in synthesising large-scale and uniform atomic layers of 2D TMDs, their quality still falls short of mechanically exfoliated flakes. Achieving single-layer controllability, defect-free film growth, TMD doping, and alloying are vital challenges that must be overcome for practical applications and improved synthesis processes.

As advanced synthesis techniques developed in the lab find practical applications, we can anticipate the emergence of 2D TMDs with currently only envisioned capabilities, revolutionising the fields of electronics, optoelectronics, and energy technology.

**Authorship Contribution Statement**

**Mitesh B. Solanki**: Conceptualization, Methodology, Writing Original Draft, Supervision. **Trilok Akhani**: Methodology, Review & Editing. **Sultan Alshehery**: Data Curation, Review

**Data Availability**

**Conflicts of Interest**

**Funding Statement**

The author has no financial relationships related to this article to disclose.

### Acknowledgement

The authors extend their appreciation to the Deanship of Scientific Research at King Khalid University, Saudi Arabia, for funding this work through the Research Group Program under Grant No: RGP 2/318/44.